\documentstyle[aps,prl]{revtex}
\begin{document}
\draft
\twocolumn[\hsize\textwidth\columnwidth\hsize\csname @twocolumnfalse\endcsname
\title{Bound states of magnons in the  S=1/2 quantum spin ladder.
}
\author{O.P. Sushkov and V.N. Kotov}
\address{School of Physics, University of New South Wales, Sydney 2052, Australia}

\maketitle

\begin{abstract}
The excitation spectrum of the  two-chain Heisenberg ladder
 with antiferromagnetic interactions  is studied.
Our approach is based on the description of the excitations
as triplets above a strong-coupling singlet ground state.
 The quasiparticle spectrum  is calculated by treating the
excitations as a dilute Bose gas with infinite on-site repulsion.
Additional singlet ($S=0$) and triplet ($S=1$)  excitations are found as 
quasiparticle bound states.  
 Thus it is demonstrated that the spectrum consists of
one elementary triplet, one composite triplet, and one composite
singlet.  
\end{abstract}

]

Quantum spin ladders  
have recently become   a subject of considerable interest,  
 mainly because of their relevance to 
a variety of quasi one-dimensional materials \cite{Dagotto}.
 The  ground state properties and the quasiparticle spectrum of the 
two-leg $S=1/2$ spin  ladder have been studied by a number of techniques, including
bosonization \cite{Shelton}, exact diagonalizations \cite{Exact},
density matrix renormalization group calculations \cite{DMRG} and  
strong-coupling expansions \cite{O,Copalan,Eder}. 
The picture emerging from these studies is that of a gapped excitation
spectrum and thus disordered ground state for any value of the 
inter-chain coupling $J_{\perp}$. 
Quite recently  a  state with energy  above the one-particle excitation  but
below the two-particle continuum was observed in
(VO)$_{2}$P$_{2}$O$_{7}$, believed to have the geometry of a ladder 
\cite{Garrett}. Subsequently this material was found to 
be better described as a dimerized chain \cite{Garrett1} which
in turn triggered theoretical work on two-particle bound states
 in this model \cite{BS}. The above work has 
 raised the important issue of 
possible low-lying two-particle excitations  in quasi one-dimensional models
with a gapped spectrum.  

In the present Letter  we develop a formalism to describe
two-particle bound states in dimerized quantum spin systems
and consider the spin ladder as an example. 
We find that a scalar (S=0) and vector (S=1) bound state 
exist below the continuum and we calculate their properties.  

Consider the Hamiltonian of  two coupled $S=1/2$ chains (spin ladder):
\begin{equation}
\label{H}
H=\sum_i\left\{J{ \bf S}_i.{\bf S}_{i+1}+
J{\bf S}_i^{\prime}.{\bf S}_{i+1}^{\prime}+
J_{\perp}{\bf S}_i.{\bf S}_i^{\prime}  \right\},
\end{equation}
where   
the  intra-chain ($J$) and the inter-chain ($J_{\perp}$) 
interactions are assumed 
antiferromagnetic $J,J_{\perp} >0$. 
At large $J_{\perp}\gg J$ the ground state  consists
 of inter-chain  spin singlets
$|GS\rangle=|1,0\rangle |2,0\rangle |3,0\rangle ...$,
where $|i,0\rangle={1\over{\sqrt{2}}}\left[|\uparrow\rangle_i
|\downarrow\rangle_i^{\prime}-|\downarrow\rangle_i
|\uparrow\rangle_i^{\prime}\right]$. Since each  singlet can be excited into
a  triplet state it is natural to introduce a
  creation operator $t_{\alpha i}^{\dag}$  for  this 
excitation: 
\begin{equation}
|i,\alpha \rangle=t_{\alpha i}^{\dag}|i,0\rangle, \ \ \alpha=x,y,z.
\end{equation}
The representation of the spin operators in terms of $t_{\alpha i}^{\dag}$
was introduced by Sachdev and Bhatt \cite{Sachdev}, and after its
application to (1) one finds   \cite{Copalan,Eder}:

\begin{eqnarray}
\label{H1}
H&=&\sum_{i,\alpha,\beta}\left\{J_{\perp}t_{\alpha i}^{\dag}t_{\alpha i}+
{J\over{2}}\left(t_{\alpha i}^{\dag}t_{\alpha i+1}+
t_{\alpha i}^{\dag}t_{\alpha i+1}^{\dag}+\mbox{h.c.}\right) \right.
\nonumber \\
& & \left. + {J\over{2}}\left(t_{\alpha i}^{\dag}
t_{\beta i+1}^{\dag}t_{\beta i}t_{\alpha i+1}
-t_{\alpha i}^{\dag}t_{\alpha i+1}^{\dag}t_{\beta i}t_{\beta i+1}\right)
\right\}.
\end{eqnarray}
In addition, the hard core constraint  
$t_{\alpha i}^{\dag}t_{\beta i}^{\dag}=0$ has to be enforced on every
site \cite{Sachdev}. This exclusion of double occupancy reflects  the quantization
of spin and ensures the uniqueness of the mapping from (1) to (3).
At the quadratic level the Hamiltonian (\ref{H1}) can be  diagonalized
by a combination of Fourier and   
 Bogoliubov transformations 
$ t_{\alpha k} = u_{k} \tilde{t}_{\alpha k}
+  v_{k} \tilde{t}_{\alpha -k}^{\dagger}$.
This gives the excitation spectrum:
$\omega_k^{2} = A_k^{2} - B_k^{2}$,
where $A_{k} = J_{\perp} + J \cos k$ and $B_{k} = J  \cos k$.

In order to address the problem of bound states of two triplets 
a reliable description of the one-particle spectrum is needed.
We find, in agreement with previous work \cite{Sachdev,Copalan},
that  the effect  of the quartic terms in (3) 
on the triplet spectrum is  small
and therefore we proceed by treating    these terms in mean field theory. 
 This is equivalent to
taking into account only   one-loop diagrams (first order in $J$).
These diagrams lead to the renormalization:
\begin{equation}
\label{AB}
A_k = J_{\perp} +J(1+2 f_1) \cos k, \ B_k = J(1-2 g_1)\cos k,
\end{equation}
where
\mbox{$f_1= < t_{\alpha i}^{\dagger}t_{\alpha i+1}> =
N^{-1}\sum_q v_q^{2} \cos q $} and
\mbox{$g_1= < t_{\alpha i} t_{\alpha i}> =
N^{-1}\sum_q  u_q v_q \cos q $}.
 
The dominant contribution to the spectrum renormalization 
is related to  the hard core condition. Previous  treatments
have used    mean-field  approximations  \cite{Copalan,Sachdev}  and are
essentially uncontrolled, especially for a quasi-1D system. 
 To deal with the constraint we will use the diagrammatic approach      
 developed by us in Ref.\cite{K}.
An infinite on-site repulsion is introduced in this approach in order
to forbid double occupancy: 
\begin{equation} 
\label{U}
H_{U} = \frac{U}{2} \sum_{i,\alpha \beta}
t_{\alpha i}^{\dagger}t_{\beta i}^{\dagger}t_{\beta i}t_{\alpha i},
 \ \ U \rightarrow \infty
\end{equation}
Since the interaction is infinite, the exact 
scattering amplitude  $\Gamma_{\alpha\beta,\gamma\delta}(K) =
\Gamma(K) \delta_{\alpha\gamma}\delta_{\beta\delta}$,
$K\equiv (k,\omega)$, 
 for the triplets has to be found.
This quantity  satisfies the Bethe-Salpeter equation, shown
in Fig.1(a) and depends on the total energy and momentum of the incoming
particles $K = K_{1} + K_{2}$. 
The interaction (5) is local and non-retarded which allows for 
an analytic solution of
the equation for $\Gamma$:
\begin{eqnarray}
\label{Gam}
\Gamma^{-1}(K) & = & i \int \frac{ d^{2}Q}{(2\pi)^2}
G(Q)G(K - Q) =  \nonumber \\
=&-&\frac{1}{N} \sum_q \frac{u_q^{2}
u_{k- q}^{2}}{\omega - \omega_q - \omega_{k- q}}
 + \left\{ \begin{array}{c} u \rightarrow v \\
\omega \rightarrow -\omega \end{array} \right\}
\end{eqnarray}
Here $G(Q)$ is the normal Green's function (GF)  
$G(k,t)=-i<T(t_{k\alpha}(t)t_{k\alpha}^{\dagger}(0))>$
and the Bogoliubov coefficients  $u_k^{2}, v_k^{2} =
\pm 1/2 + A_k/2\omega_k$.
The basic approximation made  in the derivation of $\Gamma(K)$
is the neglect of all anomalous scattering vertices, which are
present in the  theory due to  the existence of anomalous GF's,
$G_{a}(k,t)=-i<T(t_{-k\alpha}^{\dagger}(t)t_{k\alpha}^{\dagger}(0))>$.
Our crucial observation \cite{K} is that
all   anomalous contributions are suppressed by 
a small  parameter which is present in the theory -
 the density of triplet excitations $n= \sum_{\alpha} 
<t_{\alpha i}^{\dagger}t_{\alpha i}> = 3N^{-1}\sum_{q}
v_{q}^{2}$.         
We find that  
\mbox{$n\approx 0.1 \ (J_{\perp}/J=2)$}, $n\approx 0.25 \ (J_{\perp}/J=1)$
and it generally increases as $J_{\perp}$ decreases.
Since  summation of ladders with anomalous  GF's
brings additional powers of $v_{q}$ into $\Gamma$, their contribution is small
compared to the dominant one of Eq.(6). For consistency we also neglect
the second term in Eq.(6).
 Thus the triplet excitations
can be viewed as a strongly-interacting dilute Bose gas and Eq.(6)
as the first term in an expansion in powers of the gas parameter $n$
\cite{K}.      

The self-energy, corresponding to the scattering amplitude $\Gamma$ is given 
by the diagrams in Fig.1(b):
\begin{equation}
\label{Sig}
\Sigma(k,\omega) = \frac{4}{N}\sum_{q} v_q^{2}
\Gamma(k + q, \omega - \omega_q). 
\end{equation}
In order to find the renormalized spectrum, one has to solve
the coupled Dyson's equations for the normal and anomalous GF's. 
After separating the result for the normal GF  
 into a quasiparticle contribution and
incoherent background, we find \cite{K}:
\begin{equation}
\label{Gnn}
G(k,\omega) = \frac{Z_k U_k^{2}}{\omega - \Omega_k+i\delta} - 
\frac{Z_kV_k^{2}}{\omega + \Omega_k-i\delta} + G_{inc}. 
\end{equation}
The renormalized triplet spectrum  and the renormalization constant are:
\begin{eqnarray}
\label{OZ}
\Omega_k &=& Z_k \sqrt{[ A_k + \Sigma(k,0)]^{2}- B_k^{2}},\\
Z_k^{-1} &=& 1 - \left(\frac{\partial \Sigma}{\partial\omega}
\right)_{\omega =0}\nonumber .
\end{eqnarray}
The renormalized Bogoliubov coefficients in (\ref{Gnn}) are:
\begin{equation}
\label{UV}
U_k^{2},V_k^{2} = \pm \frac{1}{2} +
\frac{ Z_k[ A_k +\Sigma(k,0)]}{2\Omega_k}.
\end{equation}
Equations (\ref{Gam},\ref{Sig},\ref{OZ},\ref{UV}) have to be solved 
self-consistently for $\Sigma(k,0)$ and $Z_k$. From Eq.(\ref{Gnn}) it  
also follows  that one has to replace $u_k \rightarrow \sqrt{Z_k} U_k, \ 
v_k \rightarrow \sqrt{Z_k} V_k$
in (4), (\ref{Gam}) and (\ref{Sig}) (see also Eq.(14) below).

Let us demonstrate how this approach works in the strong-coupling
limit  $J_{\perp} \gg J$.
To first order in $J/J_{\perp}$,  $A_k=J_{\perp}+ J \cos k$ and 
$B_k=J\cos k$. This leads to $\omega_k \approx A_k = J_{\perp}+J\cos k$,
$u_k \approx 1$, $v_k \approx - (J/2J_{\perp})\cos k$
and $f_1=0$, $g_1=-J/4J_{\perp}$. Substitution 
 into (4), (\ref{Gam}) and (\ref{Sig}) gives 
\begin{eqnarray}
\label{1j}
&\Gamma(k,\omega)&=2J_{\perp}-\omega, \\
&\Sigma(k,\omega)&= (J/J_{\perp})^{2}(3J_{\perp}-\omega)/2.\nonumber
\end{eqnarray}
Then from Eq.(\ref{OZ}) we find  the quasiparticle
 residue $Z=1-(1/2)(J/J_{\perp})^2$ and
the dispersion
\begin{equation}
\label{o}
\frac{\Omega_k}{J}=\frac{J_{\perp}}{J}
+ \cos k +{{3 J }\over{4 J_{\perp}}}-
{{J}\over{4 J_{\perp}}}\cos 2k.
\end{equation}
The result (12) agrees with that obtained by direct $J/J_{\perp}$
expansion \cite{2} to this order. 
For arbitrary $J_{\perp}$ a self-consistent numerical solution of
 Eqs.(\ref{Gam},\ref{Sig},\ref{OZ},\ref{UV}) is required.
 The triplet excitation spectrum  obtained
from this solution for $J_{\perp}/J=2$ 
is shown in Fig.3. For comparison the dispersion  obtained by  8-th order
 dimer series expansion \cite{O} is also plotted. The agreement between
the two curves is excellent which reflects the smallness of the
triplet density $n\approx0.1$. However for smaller values of the inter-chain
 coupling, 
e.g. $J_{\perp}=J$ the disagreement between our and the numerical results 
is as large as 20\% at $k=0$ which reflects a  transition into
a regime, dominated by very strong quantum fluctuations. Since a   more refined
analysis is required in this regime  from now on we will concentrate on  
the case $J_{\perp}/J=2$.

The quartic interaction in the
Hamiltonian (\ref{H1}) leads to attraction between two triplet excitations.
We will show that the attraction is strong enough to form a singlet
(S=0) and a triplet (S=1) bound state.
Let us consider the 
scattering of two triplets:
$q_1\alpha +q_2\beta \to q_3\gamma+q_4\delta$ 
and  introduce the total ($Q$) and relative ($q$) momentum of the pair    
$q_1=Q/2+q$, $q_2=Q/2-q$, $q_3=Q/2+p$, and $q_4=Q/2-p$.
The bare (Born) scattering amplitude is (see Fig.2(a)):
\begin{eqnarray}
\label{M1}
M_{\alpha\beta,\gamma\delta}&=&
J \left(\delta_{\alpha\gamma}\delta_{\beta\delta}
 - \delta_{\alpha\beta}\delta_{\gamma\delta}\right)\cos(q+p)+ \nonumber \\
&& J \left( \delta_{\alpha\delta}\delta_{\beta\gamma} 
- \delta_{\alpha\beta}\delta_{\gamma\delta}\right)\cos(q-p) + \nonumber \\
&&U(\delta_{\alpha\gamma}\delta_{\beta\delta} + 
\delta_{\alpha\delta}\delta_{\beta\gamma}). 
\end{eqnarray}
The $J$ and the $U$ terms arise  from the 
quartic interaction in (3) and the constraint
(\ref{U}) respectively.
We also have to take into account that the triplet excitation differs
from the bare one due to the Bogoliubov transformation and the quasiparticle
residue. Therefore we have to make the substitution:
\begin{equation}
\label{Ms1}
M_{\alpha\beta,\gamma\delta} \to \sqrt{Z_{q_1}} U_{q_1}
      \sqrt{Z_{q_2}} U_{q_2}
      \sqrt{Z_{q_3}} U_{q_3}
      \sqrt{Z_{q_4}} U_{q_4} M_{\alpha\beta,\gamma\delta}.
\end{equation}
The  bound state satisfies the  Bethe-Salpeter equation for the  
poles of the exact scattering amplitude $\tilde{M}$. 
 This equation is presented graphically in Fig.2(b) and has the form
\cite{remark}:    
\begin{equation}
\label{BS}
\left[E_Q-\Omega_{Q/2+q}-\Omega_{Q/2-q}\right]\psi(q)=
 \frac{1}{2}
\int {{dp}\over{2\pi}}M(Q,q,p)\psi(p).
\end{equation}
Here $M(Q,q,p)$ is the scattering amplitude in the appropriate
channel (see Eqs.(16,19) below), 
 $E_Q$ is the energy of the bound state and $\psi(q)$ is the
two-particle wave
function. Let us introduce the  minimum energy for two 
excitations with given total momentum (lower edge of the two-particle
continuum) 
$E_Q^c= \mbox{min}_q \left\{\Omega_{Q/2+q}+\Omega_{Q/2-q}\right\}$.
If  a bound state exists then its energy is lower
than the continuum $E_Q < E_Q^c$. The binding energy is defined as  
$\epsilon_Q=E_Q^c-E_Q > 0$.

In the singlet (S=0) channel 
 the scattering amplitude  is:
\begin{equation}
\label{Ms}
M^{(0)}={1\over{3}}\delta_{\alpha\beta}\delta_{\gamma\delta}
M_{\alpha\beta,\gamma\delta}=-4 J \cos q \cos p +2U.
\end{equation}
First we consider the strong-coupling limit    
 $J_{\perp} \gg J $. In this limit $Z_q=U_q=1$ and 
$\Omega_q=J_{\perp}+J\cos q$. The edge of the continuum   is 
$E_Q^c=2J_{\perp}-2 J|\cos Q/2|$ and therefore 
 the Bethe-Salpeter equation (\ref{BS})
is reduced to the simple form
\begin{eqnarray}
\label{BS1}
\lefteqn{
\left[\epsilon_Q^{(0)}+2JC_Q(1+\cos q)\right]\psi(q)= } \nonumber \\
 & &= 2J\cos q\int {{dp}\over{2\pi}} \cos p \psi(p)
-U\int {{dp}\over{2\pi}} \psi(p).
\end{eqnarray}
We have introduced here the notation $C_Q=|\cos Q/2|$.
The solution of (17) is:
\begin{eqnarray}
\label{res}
\psi^{(0)}(q) &=& \sqrt{2(1-C_Q^{2})}
\frac{\cos q +C_Q}{\epsilon_Q^{(0)}/J+2C_Q(1-\cos q)}\\
\epsilon_Q^{(0)}&=&J(1-C_Q)^2. \nonumber 
\end{eqnarray}
We stress that the infinite on-site repulsion enforces the condition
$\int dp \psi(p)=0$ which means that the bound state is d-wave like.
Thus we see that in the strong-coupling limit a singlet bound state
always exists. At $J_{\perp}=2J$ Eq.(15) with the substitutions
(16,14) has to be solved
numerically and the result is presented in Fig.3. We find that
for  $k \lesssim 2\pi/5$ the binding energy is practically zero in this
case.

In the triplet (S=1) channel   the  scattering amplitude is:
\begin{equation}
\label{Mv}
M^{(1)}={1\over{2}}\epsilon_{\mu\alpha\beta}\epsilon_{\mu\gamma\delta}
M_{\alpha\beta,\gamma\delta}=-2J\sin q \sin p.
\end{equation}
In this formula there is no summation over the index  $\mu$ which gives
the spin of the bound state.
By solving the Bethe-Salpeter equation in the limit $J_{\perp} \gg J$ 
we obtain for the wave-function and the binding energy:
\begin{eqnarray}
\label{res1}
\psi^{(1)}(q) &=&\sqrt{1/2-2C_Q^{2}}
\frac{\sin q}{\epsilon_Q^{(1)}/J+2C_Q(1+\cos q)}\\
\epsilon_Q^{(1)}&=&2J(1/2-C_Q)^2, \ \  C_Q < 1/2. \nonumber
\end{eqnarray}
For $C_Q >1/2$ we find $\epsilon_Q^{(1)}=0$ which means that the triplet bound
state only exists for momenta $k>Q_{c}=2\pi/3$ (in the strong-coupling limit).
 At $J_{\perp}=2J$ the numerical solution of Eq.(15), plotted in Fig.3. shows that
the bound state exists down to $k\approx \pi/2$.
We  find that the coupling in the triplet channel is  
weaker than the one  in the singlet channel.

The size of the bound state is determined by the spatial
extent of the bound state wave function:

\begin{equation}
R_{rms} = \sqrt{<r^{2}>} = \left\{ \int \left( \frac{\partial \psi(q)}{\partial q}
\right)^{2} 
\frac{dq}{2\pi} \right\}^{1/2}
\end{equation}
In the strong-coupling limit we find, by substituting Eqs.(18,20):

\begin{equation}
R_{rms} = \left\{ \begin{array}{ll}
 (1+ C_{Q}^{2})^{1/2}(1 - C_{Q}^{2})^{-1},& S=0\\
 (1+4 C_{Q}^{2})^{1/2}(1 -4C_{Q}^{2})^{-1},& S=1
\end{array}
\right. 
\end{equation}
in units of the lattice spacing.
As expected the size of the bound state increases with 
decreasing binding energy and near the threshold 
$R_{rms}\sim (\epsilon)^{-1/2}, \epsilon \rightarrow 0$, as can be seen from
Eqs.(18,20,22). The self-consistent solution for $J_{\perp}=2J$ is shown   
on Fig.4. Except for momenta, very close to the threshold, the
bound states in both channels have   typical
sizes of a few lattice spacings.

Finally, there is no bound state in the tensor ($S=2$) channel.
This can be seen from the expression for the
scattering amplitude in this case $M^{(2)} = 2J\cos q \cos p + 2U$
 which corresponds to repulsion.

In conclusion, we have studied two-particle bound states in the two-leg
antiferromagnetic Heisenberg ladder and have shown that
a singlet bound state is always present, while a triplet one
 exists only in a limited range of momenta.
In order to find the one-particle spectrum we have used a new
diagrammatic  method to take into account  the hard-core constraint  by
treating the excitations as a dilute Bose gas.
 Our approach is very
general and applicable to any spin model for which the excitations
can be described as
triplets above a strong-coupling singlet ground state.
We can claim that bound states are present practically in any system
with dimerization. By using the techniques described in this Letter
we have also proven the existence of bound states in the two-layer Heisenberg model.
 Our general  picture for the  structure of the excitation spectrum  
 fits very well into the recently suggested strong
analogy between quantum chromodynamics and quantum antiferromagnetism 
\cite{Bob}. One can consider the elementary on-site spin 1/2 as a
quark and the set of triplet and singlet excitations as vector and scalar mesons.
From this analogy one can expect that there are even more complex
excitations, such as many-particle bound states, 
 in quantum spin models of this type \cite{us}.

We would like to thank M. Kuchiev, J. Oitmaa and Z. Weihong for stimulating
discussions and J. Oitmaa for   critical reading of the manuscript.
This work was supported by the Australian Research Council.

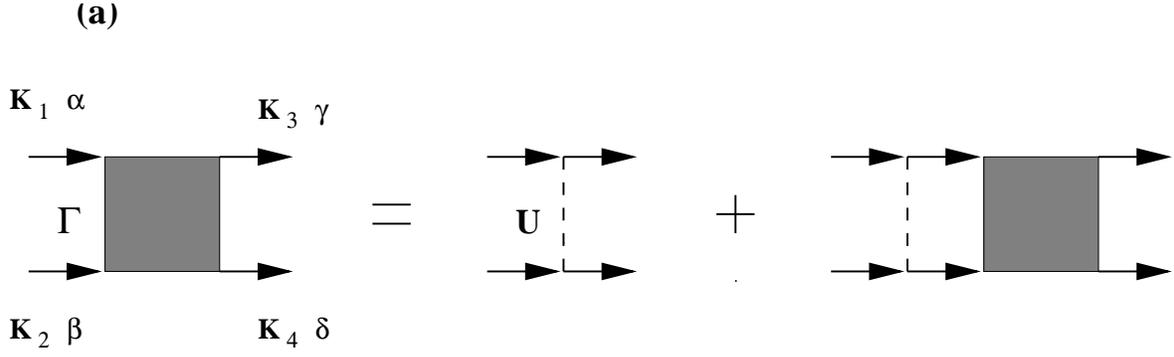
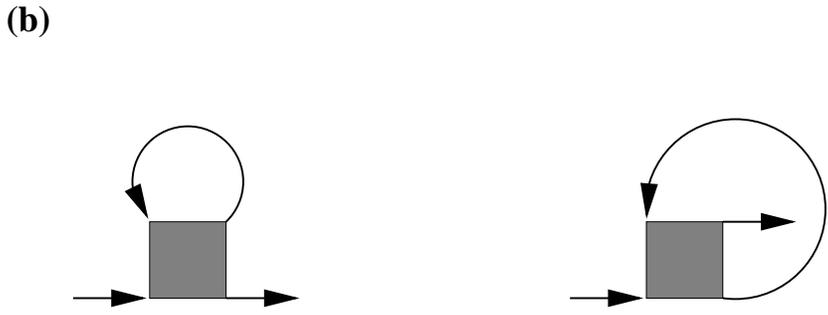
\begin{figure}
\caption
{(a) Equation for the scattering amplitude $\Gamma$.
 (b) Diagrams for the self-energy, corresponding to $\Gamma$.}
\label{fig.1}
\end{figure}

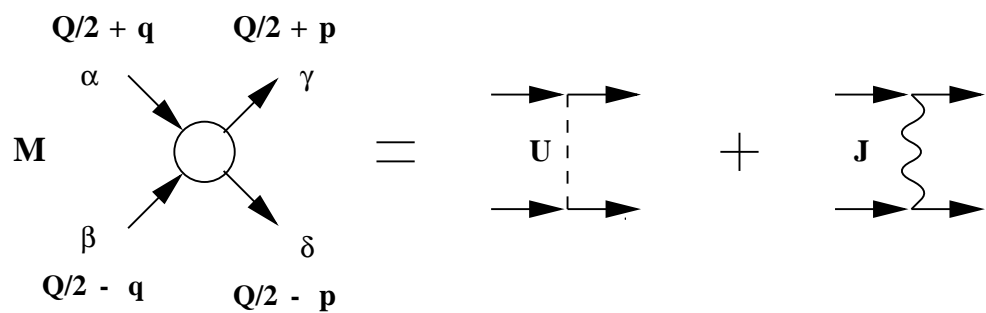
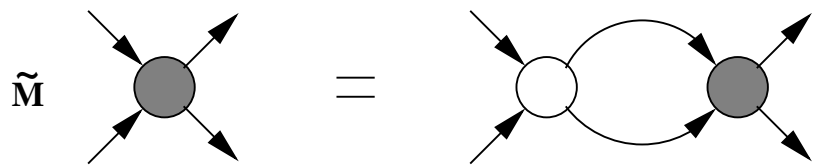
\begin{figure}
\caption
{(a) The bare (Born) scattering amplitude $M$.
 (b) The Bethe-Salpeter equation for the poles of the
exact scattering amplitude $\tilde{M}$.}
\label{fig.2}
\end{figure}
 
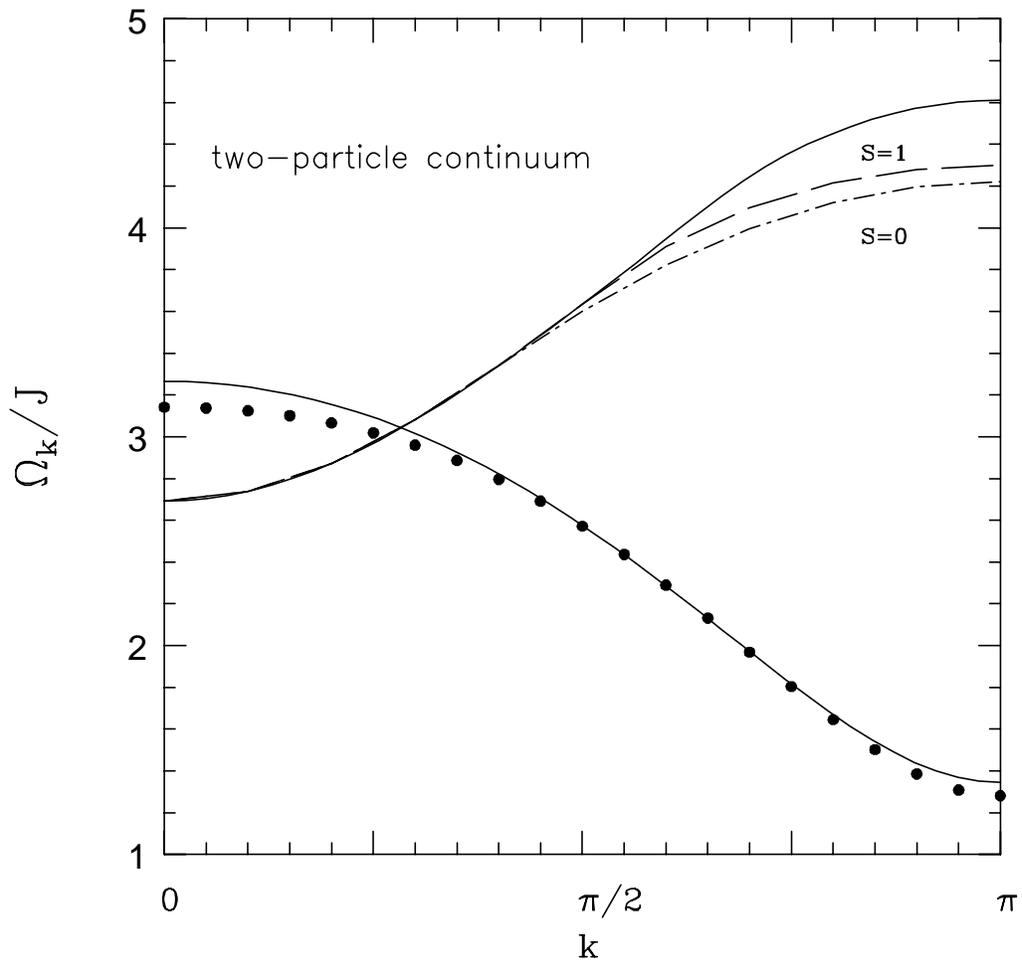
\begin{figure}
\caption
{The excitation spectrum of the ladder for $J_{\perp}=2J$.
The solid lines are the elementary triplet spectrum
(lower curve at $k=\pi$) and the
lower edge of the two-triplet continuum (upper curve).
The dots are numerical results obtained by  
dimer series expansions [5].
The dashed and the dot-dashed lines represent respectively the triplet and
singlet bound states.}  
\label{fig.3}
\end{figure}

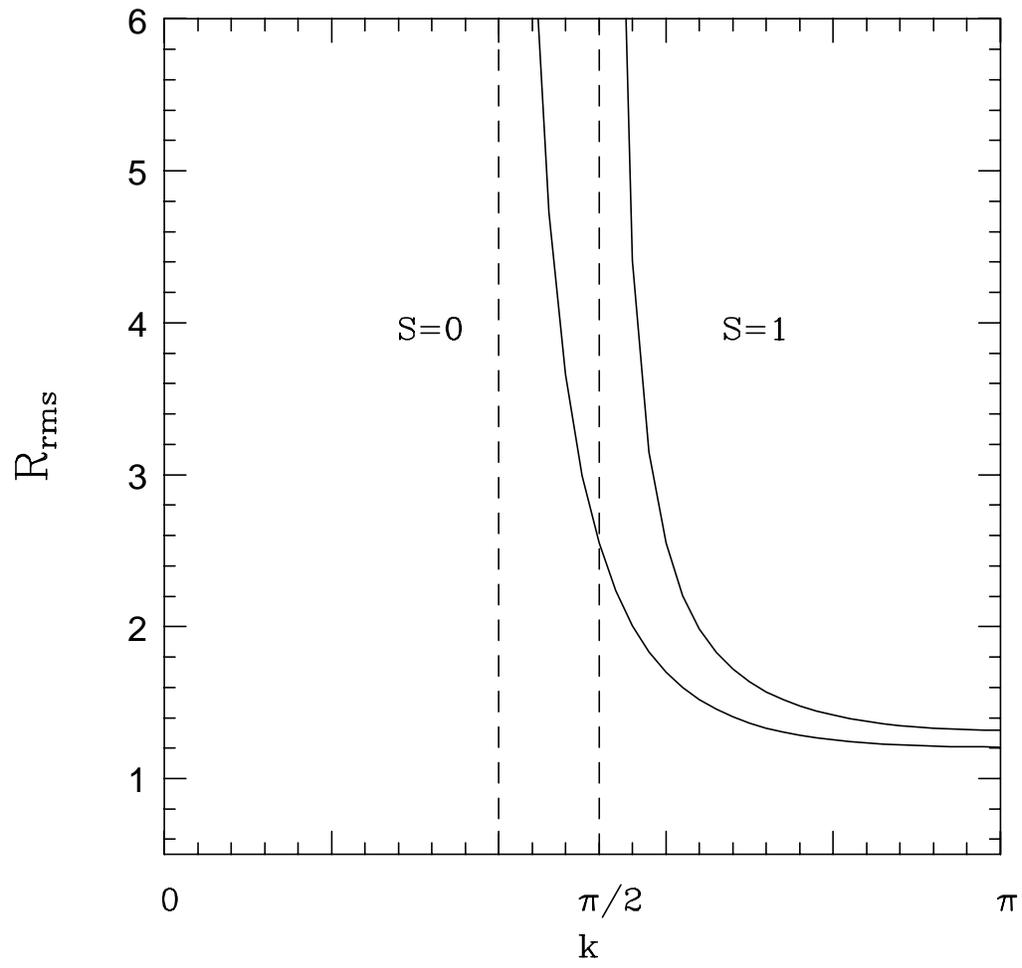
\begin{figure}
\caption
{The size $R_{rms}$ of the bound states. The vertical dashed lines
represent the points where the binding energy vanishes.}
\label{fig.4}
\end{figure}


\begin{references}
\bibitem{remark} Equation (15) can also be viewed as a two-particle
Schr\"{o}dinger equation in momentum space $H\Psi = E_{Q}\Psi$, with
$\Psi = \sum_{q} \psi^{(0)}(q)t_{\alpha q_{1}}^{\dagger}
t_{\alpha q_{2}}^{\dagger}|0>$ for the singlet and
$\Psi_{\alpha} = \sum_{q} \psi^{(1)}_{\alpha}(q)\epsilon_{\alpha \beta \gamma}
t_{\beta q_{1}}^{\dagger}t_{\gamma q_{2}}^{\dagger}|0>$ for
the triplet bound state with polarization $\alpha$. We prefer however
the more general approach presented in the text.
\end{references}
\end{document}